\documentclass[%
 reprint,
superscriptaddress,
 amsmath,amssymb,
 aps,
]{revtex4-1}

\usepackage{graphicx}
\usepackage{dcolumn}
\usepackage{bm}

\begin{document}

\preprint{APS/123-QED}

\title{Universal deformation of soft substrates near a contact line and the direct measurement of solid surface stresses}

\author{Robert W. Style}
\affiliation{%
Yale University, New Haven, CT 06520, USA 
}%

\author{Yonglu Che}%
\affiliation{%
Yale University, New Haven, CT 06520, USA
}

\author{J.S. Wettlaufer}%
\affiliation{%
Yale University, New Haven, CT 06520, USA 
}
\affiliation{NORDITA, Roslagstullsbacken 23, 10691 Stockholm, Sweden}

\author{Larry Wilen}%
\affiliation{%
Unilever Research and Development, Trumbull, CT 06611, USA
}

\author{Eric R. Dufresne}%
\email[]{eric.dufresne@yale.edu}
\affiliation{%
Yale University, New Haven, CT 06520, USA 
}
%

%

\date{\today}

\begin{abstract}
Droplets deform soft substrates near their contact lines. Using confocal microscopy, we measure the deformation of silicone gel substrates due to glycerol and fluorinated-oil droplets for a range of droplet radii and substrate thicknesses. For all droplets, the substrate deformation takes a universal shape close to the contact line that depends on liquid composition, but is independent of droplet size and substrate thickness. This shape is determined by a balance of interfacial tensions at the contact line and provides a novel method for direct determination of the surface stresses of soft substrates.
Moreover,  we measure the change in contact angle with droplet radius and show that Young's law fails for small droplets when their radii approach an elastocapillary length scale. 
For larger droplets the macroscopic contact angle is constant, consistent with Young's law.
\end{abstract}

\pacs{Valid PACS appear here}
\maketitle


Surface tension is widely known to be important for many fluid phenomena \cite{dege10}. However, solids also have tensile forces at their surface, known as surface stresses. The effects of surface stresses are particularly pronounced in thin films \cite{camm94b,spae00}, and in soft solids, where they can drive capillary waves \cite{monr98} and surface instabilities \cite{mora10,mora11}.  In particular, surface stresses are known to place fundamental limits on the resolution of microfabricated structures in soft solids \cite{hui02,jago12}

De Gennes {\emph {et al.}} \cite{dege10} noted that measurement of surface stresses in solids is ``generally perceived as an impossible task", as the effects of surface stresses are typically masked by elasticity. However, a few techniques for measuring surface stresses do exist \cite{camm94,dege10}. One approach involves measuring the bending of a microcantilever prepared with different surface properties on each of its sides. With a knowledge of the bulk elastic response of the plate, one can extract the difference in surface stresses across the faces \cite{rait00}. For soft materials, recent work suggests that surface stress can be measured by analyzing the smoothing of a soft, patterned substrate by capillarity \cite{jago12}, or by investigating the surface instability of a compressed material \cite{mora11}. These techniques are useful, but require a prior knowledge of the constitutive behavior of the material, and in some cases only provide relative values of surface stresses.

For solids, it is important to be aware of the distinction between surface free energy $\gamma$ and surface stresses $\Upsilon$. Surface free energy is the work done to form a unit area of surface, while surface stress is the tensile force at the surface of the solid. For liquids, $\gamma=\Upsilon$. In solids, $\gamma$ and $\Upsilon$ are related by the Shuttleworth equation, which shows that surfaces stress and surface energy are not necessarily equal \cite{shut50,camm94}. For many isotropic materials they are of similar magnitude, and the surface stresses are expected to be approximately isotropic \cite{peth57,camm94,spae00}. Current approaches for measuring solid surface energies suffer from similar drawbacks to techniques for measuring surface stresses \cite{chau91,john71,dege10}.

In this Letter, we demonstrate a new approach for directly measuring surface stresses in soft materials that does not require knowledge of the bulk elastic properties of the material. When a droplet rests upon a soft substrate, the surface tension of the droplet deforms the substrate at the three-phase contact line \cite{jeri11,peri09,vell10,marc12b}. We show that there is a microscopic region around the contact line where the shape is determined solely by the interfacial tensions. By measuring this shape, the surface stresses can be calculated when the surface tension of the partially wetting fluid is known. Finally, we  demonstrate that the macroscopic contact angle of a droplet shrinks for small droplets when their radii are comparable to an elastocapillary length.

\begin{figure*}
\centering
  \includegraphics[width=14cm]{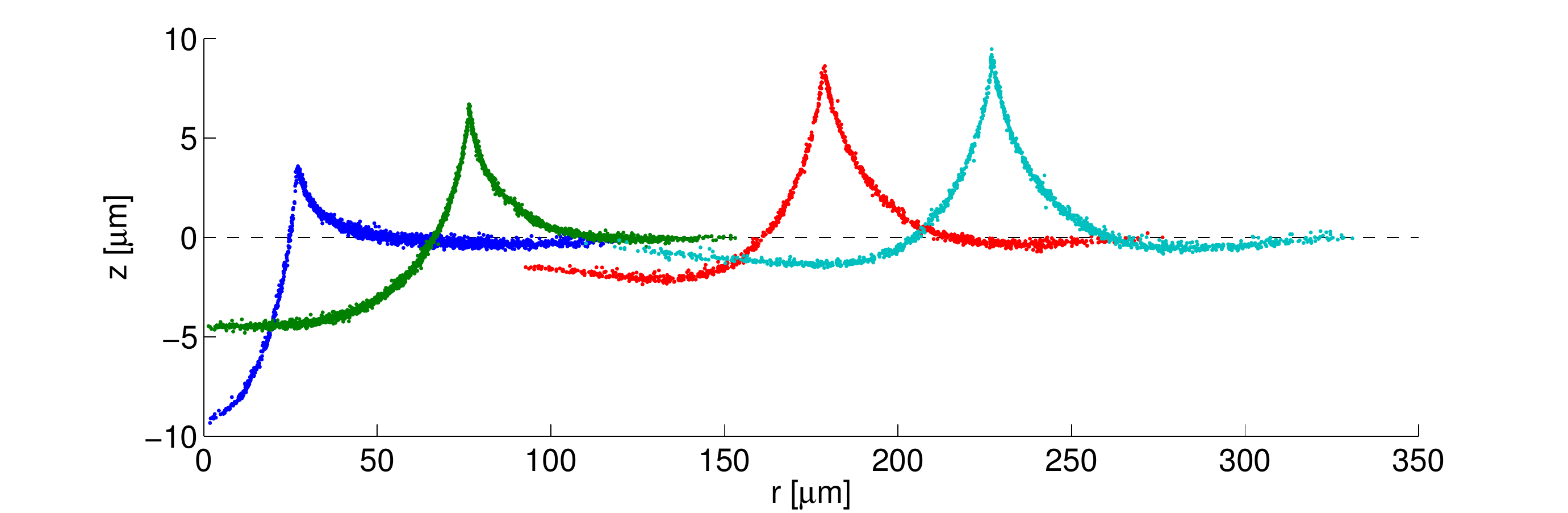}
  \caption{Surface profiles of a 50$\mu$m thick silicone gel substrate beneath partially-wetting droplets of glycerol with, from left to right, radii of 26.8, 74.5, 176.7 and 225.5 $\mu$m. The dashed line through $z=0$ corresponds to the initial surface profile before droplet deposition.}
  \label{fig:ex_profiles}
\end{figure*}

We measure the surface deformation of a soft substrate due to the presence of sessile droplets of a range of sizes. The substrates were made of a soft, elastic, silicone gel (CY52-276A/B, Dow Corning Toray), which was spin-coated into a uniform layer on a glass coverslip. From bulk rheometry, we estimated the Young's modulus of the gel as $E\approx3$kPa. For the liquid droplets, we used glycerol (Sigma-Aldrich) and fluorinated oil (Fluorinert FC-70 fluid, Hampton Research). Surface deformations were recorded by embedding fluorescent beads at the surface of the gel, and recording their positions by confocal microscopy, as described by Jerison \emph{et al.} \cite{jeri11}. There was negligible evaporation during the $\sim 20$ sec  required to image each droplet. 
Each 170$\mu$m-square field of view contained about 2000 fluorescent beads, whose three-dimensional positions were determined using Gaussian fits. We used the radial symmetry of the droplets to determine their center positions and footprint radii, $R$, by finding the values that minimized the azimuthal variation of the surface profile \footnote{See Supplemental Material at [URL will be inserted by publisher]}. Example azimuthally-collapsed profiles of glycerol droplets on a 50$\mu m$ thick substrate are shown in Figure \ref{fig:ex_profiles}. Owing to the robust radial symmetry, there is good resolution of the  surface profile at the tip of the wetting ridge.

\begin{figure}[!b]
\centering
  \includegraphics[width=9cm]{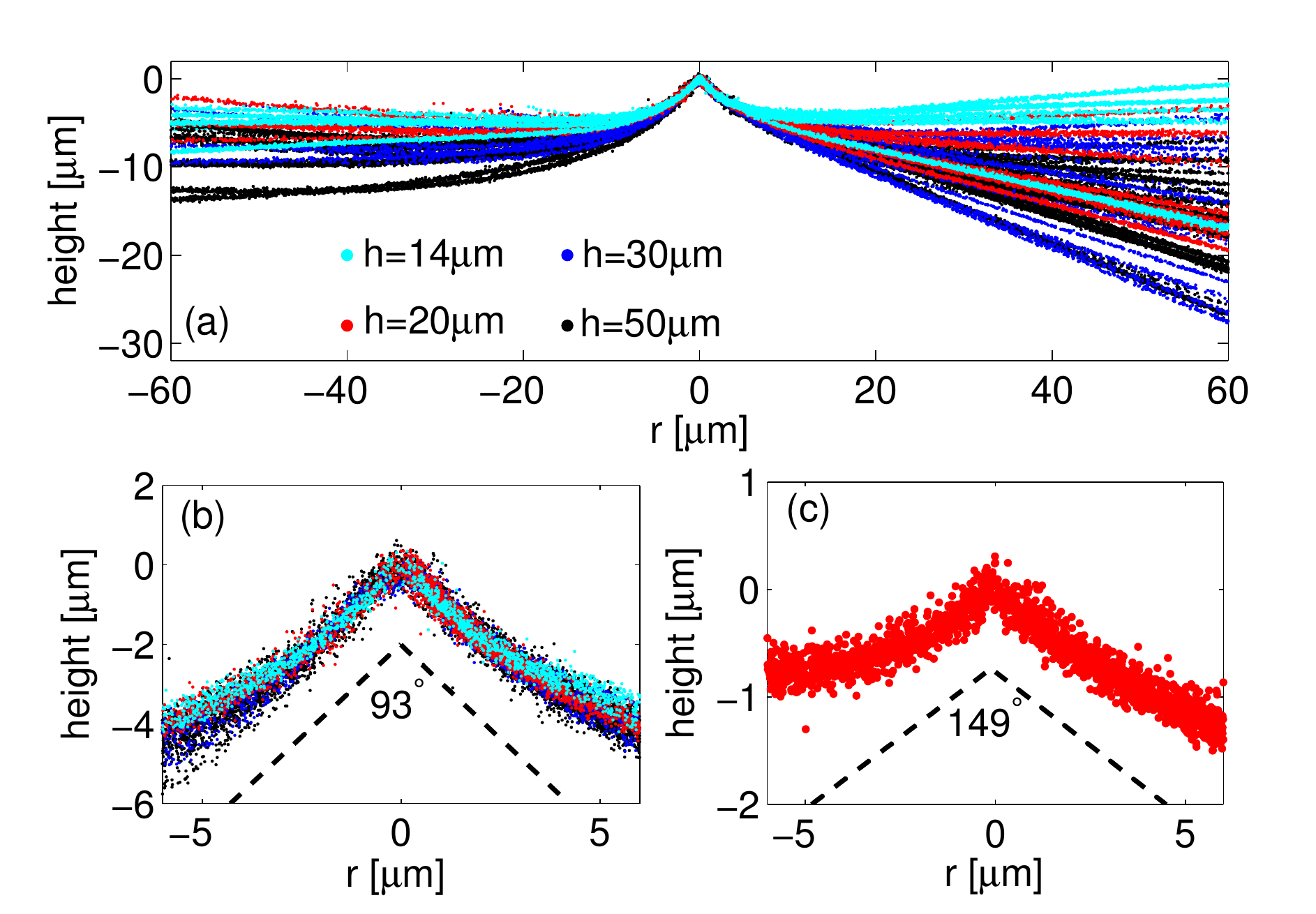}
  \caption{Universal deformation near the contact line. (a) 72 profiles of glycerol droplets with radii ranging between $20-2000\mu$m on different substrate thicknesses, shifted and rotated so that the cusp regions align. The different colors represent different substrate thicknesses, as marked in the figure. (b) A close-up of the cusp region in (a), with the dashed line showing the extracted cusp shape at the tip. (c) Close-up of the aligned cusp regions for 14 fluorinated-oil droplets with radii between 140-270$\mu$m, with the dashed line showing the extracted cusp shape.}
  \label{fig:cusp_lineup}
\end{figure}

Droplet size changes the qualitative form of the substrate deformation, as shown in Figure \ref{fig:ex_profiles}. For small droplets, the Laplace pressure is large, leading to a substantial dimple under the drop \cite{peri09}. As the droplet size increases, the pressure decreases and the dimple diminishes until the ridge is approximately symmetric \cite{jeri11,shan86}. Despite the strong variations in substrate profile with $R$, a robust feature is the locally-triangular shape of the surface at the contact line. We shall refer to this shape as a cusp.

The cusp shape appears to be universal for a given liquid/substrate pair. Figures \ref{fig:cusp_lineup}(a,b) show the surface profiles for 72 glycerol droplets of radii between $20-2000\mu m$ on 4 different substrate thicknesses ($h=$14, 20, 30 and 50 $\mu m$). Each profile is translated so that the tip of the wetting ridge is at the origin, and then rotated counterclockwise by an angle $\psi$ so that the line of symmetry of the  near-tip region is vertical. Away from the tip, there are substantial differences in profile shape. However, as seen in Figure \ref{fig:cusp_lineup}(b), all the data collapses into a sharp triangular cusp angle of (93.4$\pm 1)^\circ$ in a region within about 3$\mu m$ either side of the cusp. Figure \ref{fig:cusp_lineup}(c) shows a similar collapse of data for 14 fluorinated oil droplets with radii between $140-270\mu m$ on a 23$\mu$m thick substrate. Again the data collapses to a cusp at the contact line, this time of angle $(149.0\pm 2)^\circ$. All individual droplet profiles are provided in the supplemental material \footnote{See Supplemental Material at [URL will be inserted by publisher]}.

\begin{figure}[!h]
\centering
  \includegraphics[width=9cm]{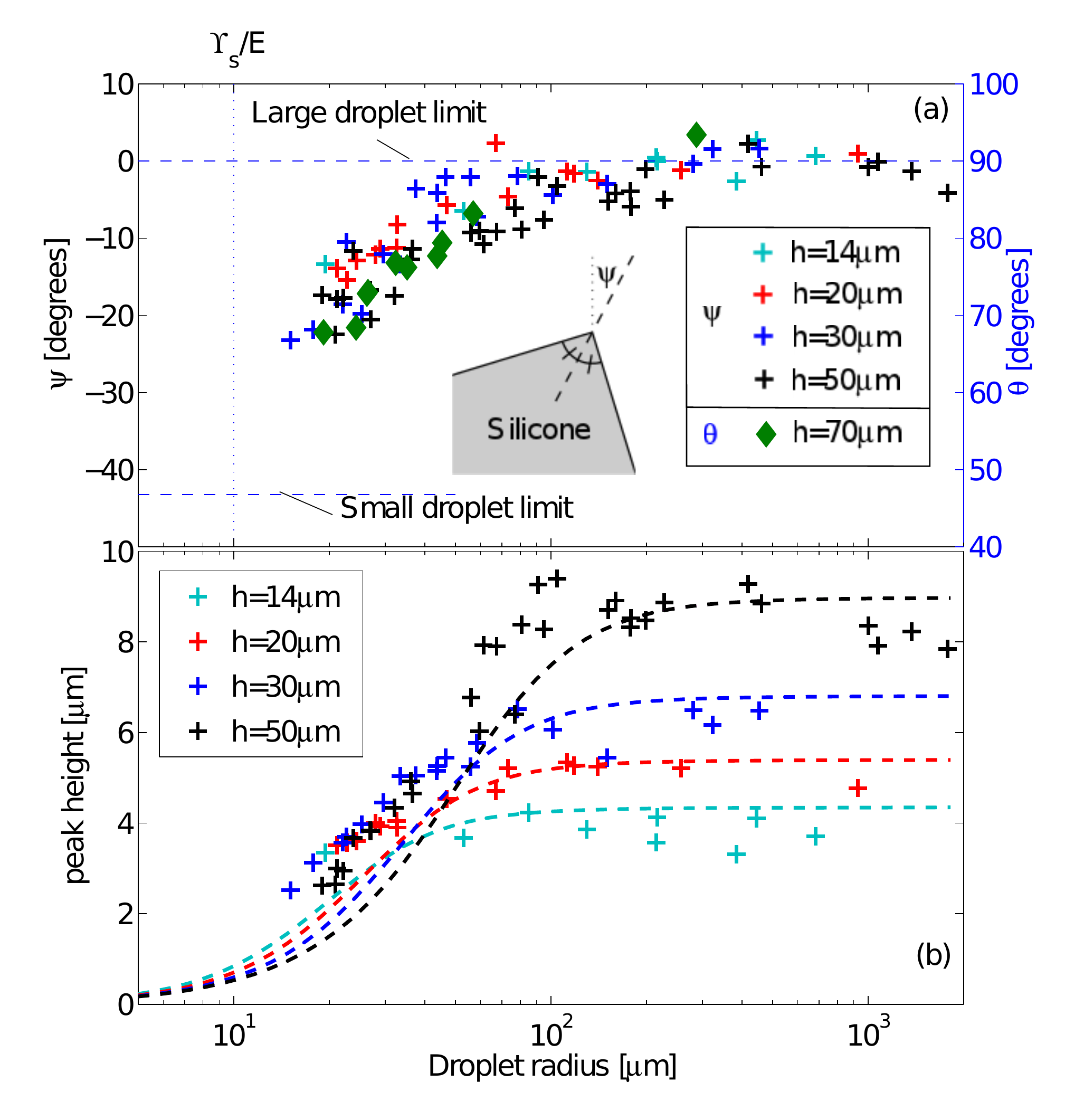}
  \caption{(a) Change in cusp orientation, $\psi$, and macroscopic contact angle, $\theta$, as a function of glycerol droplet size and substrate thickness.  Crosses and diamonds indicate $\psi$ and $\theta$ respectively. The inset shows how $\psi$ is defined.  (b) Height of wetting ridge as a function of glycerol drop size. The colors and marker shapes corresponding to substrate thicknesses are the same for both figures. Dashed curves are theoretical predictions based on measured values of the interfacial tensions, as described in the text.}
  \label{fig:angle_and_heights}
\end{figure}

While the cusp shape is universal, the cusp orientation and peak height depend on the droplet size, as shown in Figure \ref{fig:angle_and_heights}(a,b). For large droplets on all substrate thicknesses, the cusp points directly upwards ($\psi\approx 0$). As the drop size reduces towards a length scale of order 100$\mu m$, the cusp starts to rotate towards the droplet center, as can be seen in Figures \ref{fig:ex_profiles} and \ref{fig:angle_and_heights}(a). Figure \ref{fig:angle_and_heights}(b) shows the height of the wetting ridge as a function of droplet radius. For large droplets, this height approaches a constant value that depends upon the thickness of the substrate. For smaller droplets, with $R/h\lesssim 2$, the height appears to be independent of substrate thickness, depending only on $R$. The cusp shape, orientation and height now help to reveal the physical processes at work at the three-phase contact line. 

The cusp shape seen in our experiments is strikingly similar to the behavior of a three phase contact line between liquids, where contact-line geometry is entirely determined by force balance between the three surface tensions \cite{dege10}. We recently argued theoretically that the shape of a solid substrate near the tip of a wetting ridge is similarly determined by interfacial tensions, independent of bulk elasticity  \cite{styl12}. Briefly, this is because for surface perturbations of wavenumber $k$, the elastic restoring force $\sim Ek$ while the capillary restoring force $\sim \Upsilon k^2$. These are comparable when $1/k\sim\Upsilon/E$, and  capillarity dominates elasticity for features sizes $\ll O(\Upsilon_{sl}/E,\Upsilon_{sv}/E)$. Thus capillarity dominates wetting-ridge shape near the contact line. Note that this does not mean that there is liquid or plastic behavior at the contact line -- the substrate remains elastic. In this region, where bulk elasticity has no significant contribution, force balance requires that the surfaces must intersect at fixed orientations determined by the interfacial tensions, satisfying Neumann's triangle. This prediction implies that as the cusp rotates with reducing droplet size, the angle of the liquid-vapor interface -- the macroscopic contact angle, $\theta$ \footnote{We define $\theta$ as the angle from horizontal of the liquid-vapor interface at the tip of the wetting ridge} -- will rotate by the same amount.

To test the prediction that the orientation of the liquid-vapor interface is fixed relative to the cusp at the contact line, we determined $\theta$ for different-sized glycerol droplets using surface profilometry (laser profilometer with white-light probe sensor, Solarius Inc.). We measured the droplet footprint radii and their height above the undeformed substrate surface. Assuming the droplets are spherical caps, we extracted the angle at which the liquid-vapor interface would intersect the undeformed silicone substrate. On a rigid substrate with no wetting ridge, this value is the contact angle $\theta$. In our experiments, this value  systematically overestimates $\theta$ by an amount on the order the ratio of the ridge height divided by the droplet radius. This systematic error is $\approx 5^\circ$  for the smallest droplets and decreases with increasing $R$. The results for $\theta$ are displayed as diamonds in figure \ref{fig:angle_and_heights}(a), showing that $\theta$ and $\psi$ vary together with droplet size. Thus, the angles that the interfaces intersect at the contact line are fixed, in agreement with Neumann's triangle.

\begin{figure}
\centering
\includegraphics[width=8cm]{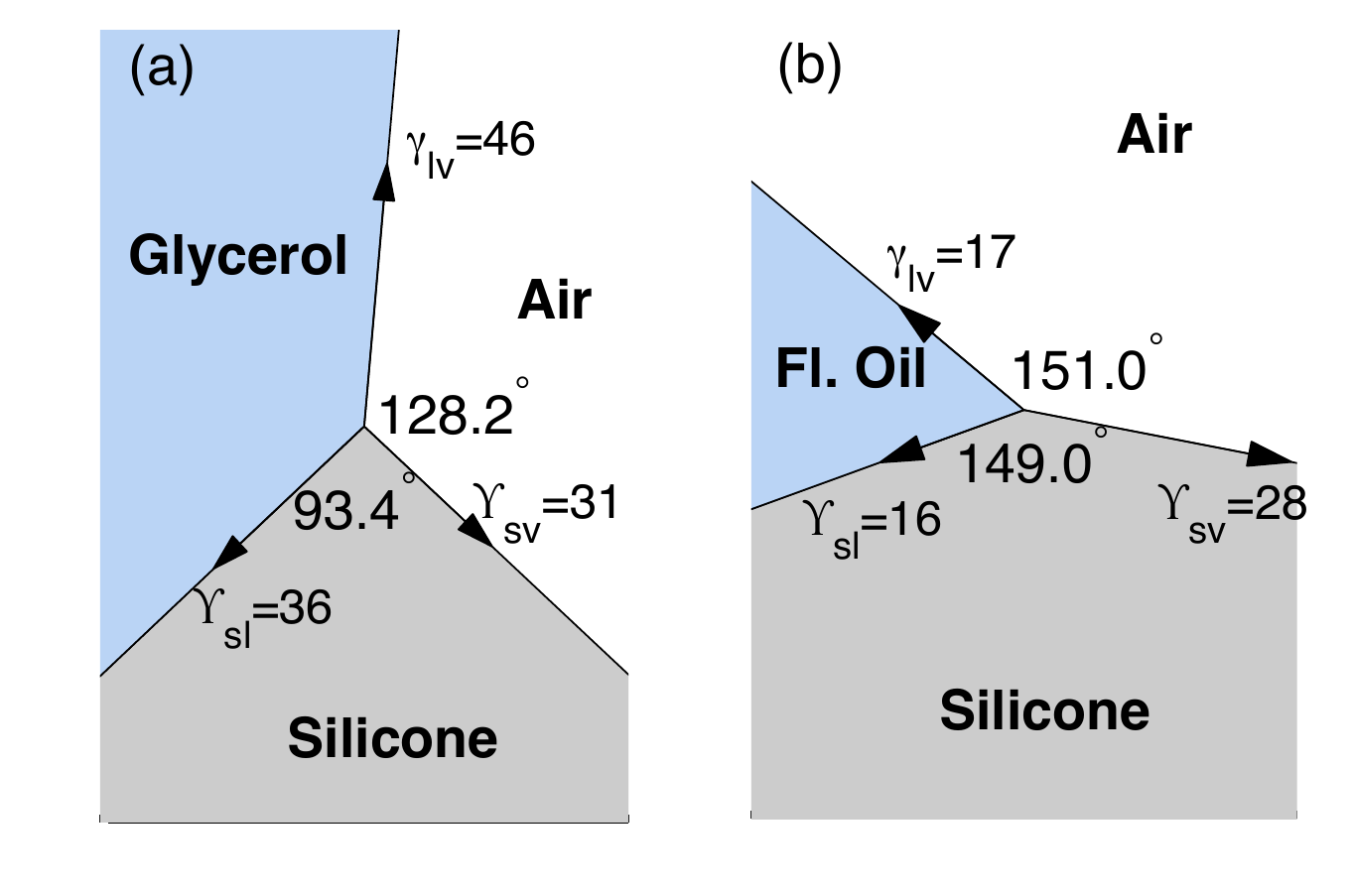}
  \caption{(a,b) The geometry of the cusp region for glycerol and fluorinated-oil droplets respectively. Measured interfacial tensions are shown as vectors with magnitude given in mN/m.}
  \label{fig:fig4}
\end{figure}

These observations suggest a new approach to measure absolute solid surface stresses from the substrate deformation near the contact line: if we know the value of any one of the interfacial tensions, and the angles between the surfaces, we can calculate the values of the remaining two tensions from the requirement of local force balance.
The surface tensions of the wetting fluids are readily-determined using the hanging droplet technique \cite{dege10}.
For fluorinated oil and pure glycerol, we found $\gamma^f_{lv}=(17 \pm 1)$mN/m and $\gamma^g_{lv}=(61\pm 1)$mN/m respectively. However, contact with the silicone substrate significantly reduced the surface tension of glycerol: droplets removed from a silicone substrate had a surface tension $\gamma_{lv}^g=(46 \pm 4)$mN/m, presumably due to liquid silicone oil in the gel substrate being wicked onto the (high energy) surface of the droplet. The confocal-microscopy measurements in Figures \ref{fig:ex_profiles}-\ref{fig:cusp_lineup} precisely specify cusp orientation, and the angle between the solid-liquid and solid-vapor interfaces. However, they do not give the orientation of the liquid-vapor interface relative to the other two interfaces.  This can be  determined with the macroscopic contact angle, $\theta$, which can be readily measured for large droplets with a  standard contact-angle goniometer  (VCA Optima, AST Products).  Conveniently,  $\psi$ and $\theta$ are independent of  droplet radius for $R>250\mu$m. Thus, we can average  micro- and macroscopic experiments over  a range of droplet sizes to obtain all the angles between the interfaces at the contact line with good accuracy.
For the largest glycerol droplets ($R>1$mm), we measured an average macroscopic angle of $\theta=95^\circ$ with an advancing contact angle $\theta_a=100^\circ$ and a receding contact angle $\theta_r=90^\circ$. For fluorinated-oil droplets ($R>1$mm), $\theta=40^\circ$ with $\theta_a=45^\circ$ and $\theta_r=35^\circ$. These measurements were accurate to within $\pm1^\circ$.
The macro- and microscopic data are combined to determine the shape of the contact line region, as shown in Figure \ref{fig:fig4}(a,b).
These results yield absolute values of the surface stresses; from the the glycerol experiments we find $\Upsilon^g_{sl}=(36\pm4)$mN/m and $\Upsilon^g_{sv}=(31\pm5)$mN/m, while from the fluorinated-oil experiments we find $\Upsilon^f_{sl}=(16\pm2)$mN/m and $\Upsilon^f_{sv}=(28\pm2)$mN/m. These are reasonable values: the measured values of $\Upsilon_{sv}$ from the two sets of experiments are consistent with each other, and are not much higher than the surface tension of liquid silicone, $21$mN/m. Combining our results with a rheological characterization of the gel, we arrive at  the governing elastocapillary length for our substrates $\Upsilon_s/E\approx 10\mu$m.

While the microscopic arrangement of the three interfaces is universal, our results show that the macroscopic contact angle depends on droplet size, in contradiction with Young's law. Specifically,  $\psi$ and $\theta$  show a pronounced decrease as $R$ reduces below $200\mu$m, as shown in Figure \ref{fig:angle_and_heights}(a). Intriguingly, this indicates that Young's law is violated for small droplets on soft substrates  \footnote{The change in $\theta$ can not be attributed to contact angle hysteresis as it is significantly more than the measured contact angle hysteresis of $\pm5^\circ$.}\cite{shan87}. This behavior is in qualitative agreement with our earlier predictions for droplets with $\Upsilon_{sl}=\Upsilon_{sv}=\Upsilon_s$ \cite{styl12}. We showed that for droplets with radius $R\gg O(\Upsilon_s/E)$, $\theta$ depends on surface energies, taking the value given by Young's law. For small droplets, $R\ll O(\Upsilon_s/E)$, the geometry of the contact line depends on the surface stresses and $\theta$ reduces to a smaller value given by Neumann's triangle -- the system has the same geometry as a droplet on a liquid substrate. The contact angle varies smoothly between these two values as the drop size decreases, with the transition occurring near the elastocapillary length, $\Upsilon_s/E$.  With our optical methods, it was not possible to resolve cusps for droplets smaller than $20\mu$m in size, and so we could not confirm that $\theta$ is independent of droplet size for $R \ll \Upsilon_s/E$. However, we can calculate the expected value of $\Delta \theta$ for such small droplets. In this limit, the interface geometry is as shown in Figure \ref{fig:fig4}(a) with the cusp rotated so that the solid-vapor interface is horizontal \cite{styl12} and $\Delta \theta=-43.3^\circ$. This prediction is shown in Figure \ref{fig:angle_and_heights}(a) as a dashed line, and is consistent with the trend suggested by the data \footnote{In our previous theory \cite{styl12} we also predicted that the contact angle would initially rise from Young's law as the drop size decreased, but this is due to an error in our calculation: when calculating the free energy of a sessile droplet we assumed that, when surface tension pulls up a ridge on a soft substrate, the change in surface energy due to changes in interfacial area are negligible in comparison to the elastic energy stored in the substrate. This is incorrect}.
 
Finally, we use the surface stresses extracted from the geometry of the interfaces at the contact line to predict the global substrate deformation. We use our previous theory which gives substrate deformations for hemispherical droplets on linear elastic substrates for the special case $\Upsilon_{sv}=\Upsilon_{sl}=\Upsilon_s$ \cite{styl12}.  This is a reasonable approximation for the glycerol experiments. To calculate substrate deformations, we assume that the silicone gel is incompressible \cite{jeri11} and use the extracted experimental values for glycerol $\Upsilon_s=(\Upsilon_{sl}+\Upsilon_{sv})/2=33.5$mN/m and $\gamma_l=46$mN/m, along with measured values of $E$ and $h$. Figure \ref{fig:angle_and_heights}(b) shows the theoretical peak height of a wetting ridge for all the substrate thicknesses and droplet radii in our experiments. The full droplet profiles are compared against theoretical profiles for each droplet in the supplement \footnote{See Supplemental Material at [URL will be inserted by publisher]}. With no fitting parameters, there is good agreement between the theory and data -- despite the fact that the material behavior is expected to be nonlinear at large strains, and that the theory only strictly holds for small surface deformations. This corroborates our measured values of $\Upsilon_{sv}$ and $\Upsilon_{sl}$.

In conclusion, we have measured the deformation of soft materials under sessile droplets.  Our results suggest that in the neighborhood of the contact line, the bulk rheological behavior of the solid is unimportant, and the local shape is controlled entirely by the droplet surface tension and the substrate surface stresses. As there must be force balance at the contact line, the surface stresses can be calculated from the angles that the phase interfaces intersect, along with the surface tension of the droplet phase. Unlike previous techniques, this provides a direct method for measuring absolute values of surface stresses, without the need for knowledge of the bulk constitutive behavior of the solid. This technique should be suitable for measuring surface stresses for any material that is sufficiently soft that the shape of the wetting ridge can be accurately measured. If surface stresses depend on the strain at the wetting ridge, this approach should be able to provide surface stresses at a range of different surface strains by varying the droplet liquid and the fluid that comprises the surrounding media. Finally, we find that Young's law breaks down for small droplets on soft substrates. This suggests that it is necessary to reconsider diverse wetting phenomena whereÊ droplets Êare  not  much larger than the elastocapillary length.
 
RWS is funded by the Yale University Bateman Interdepartmental Postdoctoral Fellowship. JSW thanks the Swedish Research Council for support. Support for instrumentation was provided by NSF (DBI-0619674).  We thank David Quer\'{e}, Howard Stone and Tom Witten for helpful conversations about this work at the \'{E}cole de Physique des Houches summer school on Soft Interfaces. We also thank Elizabeth Jerison and Anand Jagota for their helpful comments, and Rostislav Boltyanskiy for rheology measurements used in this work.



%

\end{document}